\documentclass[aps,pra,floatfix,showpacs]{revtex4} %,twocolumn

\usepackage{amsmath}
\usepackage{amsfonts}
\usepackage{amssymb}
\usepackage{bm}
\usepackage{color}
\usepackage{graphicx}
\usepackage{natbib}
\usepackage{subfigure} 
\usepackage{float}\newcommand{\comment}[1]{}

\begin{document}
\title{Replicator dynamics with turnover of players}

\author{Jeppe Juul$^1$, Ardeshir Kianercy$^2$, Sebastian Bernhardsson$^3$, and Simone Pigolotti$^4$}

\affiliation{$^1$Niels Bohr Institute, Blegdamsvej 17, DK-2100, Copenhagen, Denmark, 
$^2$Information Sciences Institute, University of Southern California, Marina del Rey, CA 90292, USA,
$^3$Swedish Defence Research Agency, SE-147 25 Tumba, Sweden,
$^4$Dept. de Fisica i Eng. Nuclear, Universitat Politecnica de Catalunya
Edif. GAIA, Rambla Sant Nebridi 22, 08222 Terrassa, Barcelona, Spain} 

\date{\today}

\begin{abstract}
  We study adaptive dynamics in games where players abandon the
  population at a given rate, and are replaced by naive players
  characterized by a prior distribution over the admitted
  strategies. We demonstrate how such process leads macroscopically to
  a variant of the replicator equation, with an additional term
  accounting for player turnover. We study how Nash equilibria and the
  dynamics of the system are modified by this additional term, for
  prototypical examples such as the rock-scissor-paper game and
  different classes of two-action games played between two distinct
  populations. We conclude by showing how player turnover can account
  for non-trivial departures from Nash equilibria observed in data
  from lowest unique bid auctions.
\end{abstract}

 %Game theory = 02.50.Le, 89.75.-k = complex systems,  89.65.Gh = Economics; econophysics, financial markets, business and management

\maketitle
%%%%%%%%%%%%%%%%%%%%%%%%%%%%%%%%%%%%%%%%%%%%%%%%%%%%%%%%%%%%
\section{Introduction}
%%%%%%%%%%%%%%%%%%%%%%%%%%%%%%%%%%%%%%%%%%%%%%%%%%%%%%%%%%%%

Perhaps the most important skill for competing agents is the ability
to adapt to a changing environment by constantly assessing and
modifying their behavior. Consequently, while game theory has been
initially mostly focused on the study of equilibria
\cite{vonneumann,jnash}, the study of adaptive dynamics has acquired
more and more relevance in recent years. Various models have been put
forward to capture the learning processes of competing individuals and
the resulting evolution of the population \cite{bush1951mathematical,
  watkins1992q, kianercy2012dynamics}. A key result is that many
agent-based algorithms of adaptive dynamics allow for a simple
macroscopic description in terms of replicator equations
\cite{fisher1930theory,Hofbauer1985, nowak2006evolutionary,traulsen}.
In the replicator dynamics, a large population of individuals
participate in a game.  Let us call $x_i$ the fraction of population
playing a given strategy, where admissible strategies are labeled as
$i = \{1, 2, 3, \ldots \}$.  The time evolution of such fractions is
given by:
\begin{equation}\label{repl}
	\frac{d}{dt} x_i=x_i(\pi_i(\mathbf{x})-\bar{\pi}(\mathbf{x})),
\end{equation}

where $\pi_i(\mathbf{x})$ is the frequency-dependent payoff of
strategy $i$, and $\bar{\pi}=\sum_i x_i \pi_i$ is the average
payoff. In this setting, the initial condition $\mathbf{x}_i^0$
represent a prior distribution of strategy preferences, before the
adaptation process takes place.  Approaches based on Eq. \eqref{repl}
have proven successful in describing systems within biology as well as
economics \cite{nowak2006evolutionary, maynard_book,
  silverberg1997evolutionary}. It can be easily shown that stable
equilibria of replicator dynamics correspond to Nash equilibria, where
no individual can benefit from changing strategy unilaterally
\cite{Hofbauer1985, Nash}.

Usually, in adaptive dynamics, one has in mind a fixed population of
players that acquire experience over time, or 
biological populations, where offsprings inherit strategies from their
parents. However, one can think of a number of concrete examples where
games are played in a more open setting, with the possibility of
players to leave the game and be replaced by less experienced ones.
On the market, new companies are founded while old companies
collapse. Within companies, experienced employees retire so that young
graduates may start their career. On general grounds, one should
expect turnover to have a profound impact on the dynamics of the game
and lead to a rich phenomenlogy. In a standard adaptive dynamics all
individuals within the population have the same degree of
experience. Conversely, here each agent sees a non-trivial mixture of
players with different experience levels. While the Nash equilibrium
strategy will be optimal against very experienced players, it would
not necessarily be the most effective to exploit naive
newcomers. Therefore, one can expect adaptive dynamics in this case to
converge to equilibria being different from Nash equilibria, and being
crucially affected by the rate of turnover, which in turns determines
the steady-state structure of the population both in terms of
experience and strategies. In such framework, interesting information
can be obtained by dissecting the experience composition of each
strategy, for example to assess which experience classes are
performing better in the game in terms of payoff.

In this paper, we demonstrate how taking into account turnover of
players leads macroscopically to a novel variant of the replicator
dynamics.  We present a derivation of the replicator equation with
turnover (Eq. \ref{repl_turnov}) in Section \ref{modelsection}.
In the remainder of the paper, we apply this equation to analyze the
effect of turnover in simple evolutionary games. We begin with the
simple paradigmatic cases of the rock-paper-scissors game and the set
of two-action games played between two different populations.  In the
latter case, we show how increasing the turnover rate can lead to
abrupt change in the equilibrium state caused by bifurcations in the
corresponding dynamical system.  We conclude the paper by showing how
this approach can provide an interpretation for the observed bid
distribution in online lowest unique bid auctions.

%%%%%%%%%%%%%%%%%%%%%%%%%%%%%%%%%%%%%%%%%%%%%%%%%%%%%%%%%%%%
\section{Model}\label{modelsection}
%%%%%%%%%%%%%%%%%%%%%%%%%%%%%%%%%%%%%%%%%%%%%%%%%%%%%%%%%%%%

We aim at generalizing the replicator equations (\ref{repl}) to
situations in which agents are replaced by inexperienced individuals
at a rate $p$. Let a large population of players engage in a game with strategies
labeled $i = \{1, 2, 3, \ldots\}$. We divide the players in
experiences classes: let $n_i(\tau,t)$ be, at time $t$, the
number of  players having been in the game for a
time $\tau$ and playing strategy $i$. The normalization condition for
the $n$'s reads:

\begin{equation}
\sum_i\sum_{\tau=0}^{\infty}\ n_i(\tau,t)=N \quad \forall t  ,
\end{equation}
where $N$ is a fixed total population size.  We also define the total
fraction of players adopting strategy $i$ at time $t$,
$x_i(t)=\sum_{\tau=0}^\infty n_i(\tau,t)/N$.  As in the standard
replicator equation (\ref{repl}), we introduce the average payoff of
strategy $i$, $\pi_i(\mathbf{x}(t))$, and the average payoff across strategies
$\bar{\pi}=\sum_i x_i \pi_i$.

For the sake of simplicity, we consider a simple adaptive dynamics in
which individuals learn of alternative strategies and their average
payoff at a rate equal to the fraction of the population that plays by
this strategy.  Further, an individual playing strategy $i$ learning
of a strategy $j$ with a higher average payoff, changes to strategy
$j$ with a probability proportional to the payoff difference.
Combining these assumptions gives an overall rate of change from
strategy $i$ to strategy $j$ equal to $n_i x_j (\pi_j - \pi_i)$, if
$\pi_j > \pi_i$.
 
Furthermore, agents leave the population at rate $p$ and are replaced
by inexperienced agents. The new agents play each strategy $i$ with a
probability proportional to a given distribution $x_i^0$.

We wish to study the time development of the $n_i(\tau,t)$. The
learning dynamics encoded in the rules above reads:

\begin{equation} \label{eq:ni}
n_i(\tau+1,t+1)-n_i(\tau,t) = 
(1-p) \left( \sum_{j: \, \pi_j < \pi_i} n_j(\tau,t) x_i(t) (\pi_i - \pi_j)  
- \sum_{j: \, \pi_j > \pi_i} n_i(\tau,t) x_j(t) (\pi_j - \pi_i)  \right)   
-p n_i(\tau,t),
\end{equation}
where the first sum represents players changing to strategy $i$ from
strategies with lower payoffs, and the second sum represents players
changing from strategy $i$ to strategies with higher payoffs.
Performing a continuous time limit, the left hand side of the above
equation becomes $\partial_t n_i(\tau,t)+\partial_{\tau} n_i(\tau,t)$.
We can now integrate the continuous time version of Eq. (\ref{eq:ni})
over $\tau$ and divide by the population size $N$, to obtain a closed
evolution equation for the strategies $x_i(t)$. Let us recall that
$n_i(0,t)=\sigma N x_i^{0}$, where the proportionality constant
$\sigma$ can be determined by imposing that the final equation
preserves the normalization condition. We also assume that, due to the
effect of turnover, one has $\lim_{\tau\rightarrow\infty}
n_i(\tau,t)=0\quad \forall i,t$.  After combining the two sums, this
results in
\begin{equation}
\frac{d}{dt} x_i(t) = (1-p)  \sum_{j} x_j(t) x_i(t) (\pi_i - \pi_j)  +\sigma x_i^{0}-p x_i.
\end{equation}
Imposing the normalization results in $\sigma=p$. This also implies
that the density of players having a given experience level at steady
state is exponentially distributed, $\sum_i n_i(\tau,\infty)=Np \exp(-p\tau)$. If we furthermore use the normalization condition $ \sum_{j}  x_j  = 1$ and the definition of the average payoff $\bar{\pi}$, we get
\begin{equation}
	\frac{d}{dt} x_i =(1-p) x_i ( \pi_i - \bar{\pi} ) + p (x_i^0-x_i) .
\end{equation}
Upon rescaling time by $1-p$ and define a rescaled turnover
rate $\chi=p/(1-p)$, we finally obtain
\begin{equation} \label{repl_turnov}
	\frac{d}{dt} x_i = x_i ( \pi_i - \bar{\pi} ) + \chi (x_i^0-x_i) .
\end{equation}
Eq. (\ref{repl_turnov}) constitutes the starting point of our
analysis. It should be clear that our specific choice of adaptive
dynamics is not crucial for deriving Eq. (\ref{repl_turnov}), and that
the same macroscopic limit could be obtained for other microscopic
adaptation rules leading to the replicator equation (\ref{repl}) in
the absence of turnover. Eq. \eqref{repl_turnov} can be of course also
derived (or justified) heuristically, for example by making an analogy
with a (damped) driven dynamical system, where the prior distribution
$x^0_i$ in the right hand side acts as a forcing term. Note that
Eq. \eqref{repl_turnov} can be formally recast in replicator form:

\begin{equation}\label{repl_turnov_cast}
	\frac{d}{dt} x_i=x_i(\pi_i-\bar{\pi})+\chi(x^0_i-x_i) 
\equiv  x_i(\tilde{\pi}_i-\bar{\pi}) ,
\end{equation}

where the ``effective payoff'' of strategy $i$, $\tilde{\pi}_i$, is defined as

\begin{equation}\label{effpayoff}
	\tilde{\pi}_i=\pi_i+\chi \left(\frac{x^0_i}{x_i}-1\right).
\end{equation}

In Eq. \eqref{repl_turnov_cast} the average payoff $\bar{\pi}$ does not contain a contribution from
the second term in Eq. \eqref{effpayoff} as its average is zero,
so that $\sum_i x_i \pi_i=\sum_i x_i \tilde{\pi}_i$. The mapping
in Eq. \eqref{repl_turnov_cast} is valid only in the interior of the
simplex ${x_i>0\ \forall i, \sum_i x_i=1}$, while at the boundary the
effective payoffs diverge. This situation has some resemblance to the
case of evolutionary dynamics in the presence of mutations
\cite{Hofbauer1985, tarnita2009mutation}. The main difference is that in our case the
``mutants'' are not characterized by new, pure strategies or random strategies, but the
mixed distribution $x_i^0$ of the existing strategies.

The fact that the replicator equation with turnover
\eqref{repl_turnov} can be rewritten in replicator form
\eqref{repl_turnov_cast} implies that the two equations share several
mathematical properties. Among these is that the mean effective payoff
for the population can not decrease along any trajectory. If the
payoff function is continuous and bounded, it therefore serves as a
Lyapunov function \cite{strogatz2001nonlinear}, which guarantees that
all game dynamics either evolve to an equilibrium point or a
closed orbit. For the replicator equation without turnover, $\chi =0$,
all Nash equilibria of a game are equilibrium points of the
dynamics~\cite{Hofbauer2003}. For positive $\chi$, the equilibrium
points are generally different from the Nash equilibria, and we
call these the \emph{turnover equilibria} of the game. In the limit
$\chi \to \infty$, agents are insensitive to the rewards and the
initial strategy $x_i^0$ is the only equilibrium point.

%%%%%%%%%%%%%%%%%%%%%%%%%%%%%%%%%%%%%%%%%%%%%%%%%%%%%%%%%%%%
\section{Rock-Paper-Scissors}
%%%%%%%%%%%%%%%%%%%%%%%%%%%%%%%%%%%%%%%%%%%%%%%%%%%%%%%%%%%%
We now apply the concept of agent turnover to the three-strategy game
of rock-paper-scissors. In this well-known, biologically relevant game
\cite{Frean, Kerr02, kirkup2004antibiotic, Nahum, sinervo1996rock}, a
large population of players can choose between the strategies rock,
paper, and scissors. The average fraction of the population employing
each strategy is denoted $x_R$, $x_P$, and $x_S$, respectively.  The
players are paired randomly in each round. Since the strategies
dominate each other cyclically, their payoffs are given by
\begin{align}
	\pi_R &= x_S - x_P \\
	\pi_P &= x_R - x_S \\
	\pi_S &= x_P - x_R .
\end{align}
Rock-paper-scissors is a zero-sum game, so the average payoff in
(\ref{repl_turnov}) is always zero.  The only Nash equilibrium of the
game is when all individuals play randomly, giving them all an average
payoff of zero \cite{szabo2007evolutionary}.

When we introduce a player turnover, na\"{\i}ve players that enter the
game choose to play rock, paper, or scissors with probability
$x^0_R$, $x^0_P$, and $x^0_S$, respectively.  Gradually, each
individual change strategy according to Eq. \eqref{eq:ni}, resulting
in the mean behavior given by Eq. \eqref{repl_turnov}.

\begin{figure}[tb]
	\begin{center}
	\includegraphics[width=8.4cm]{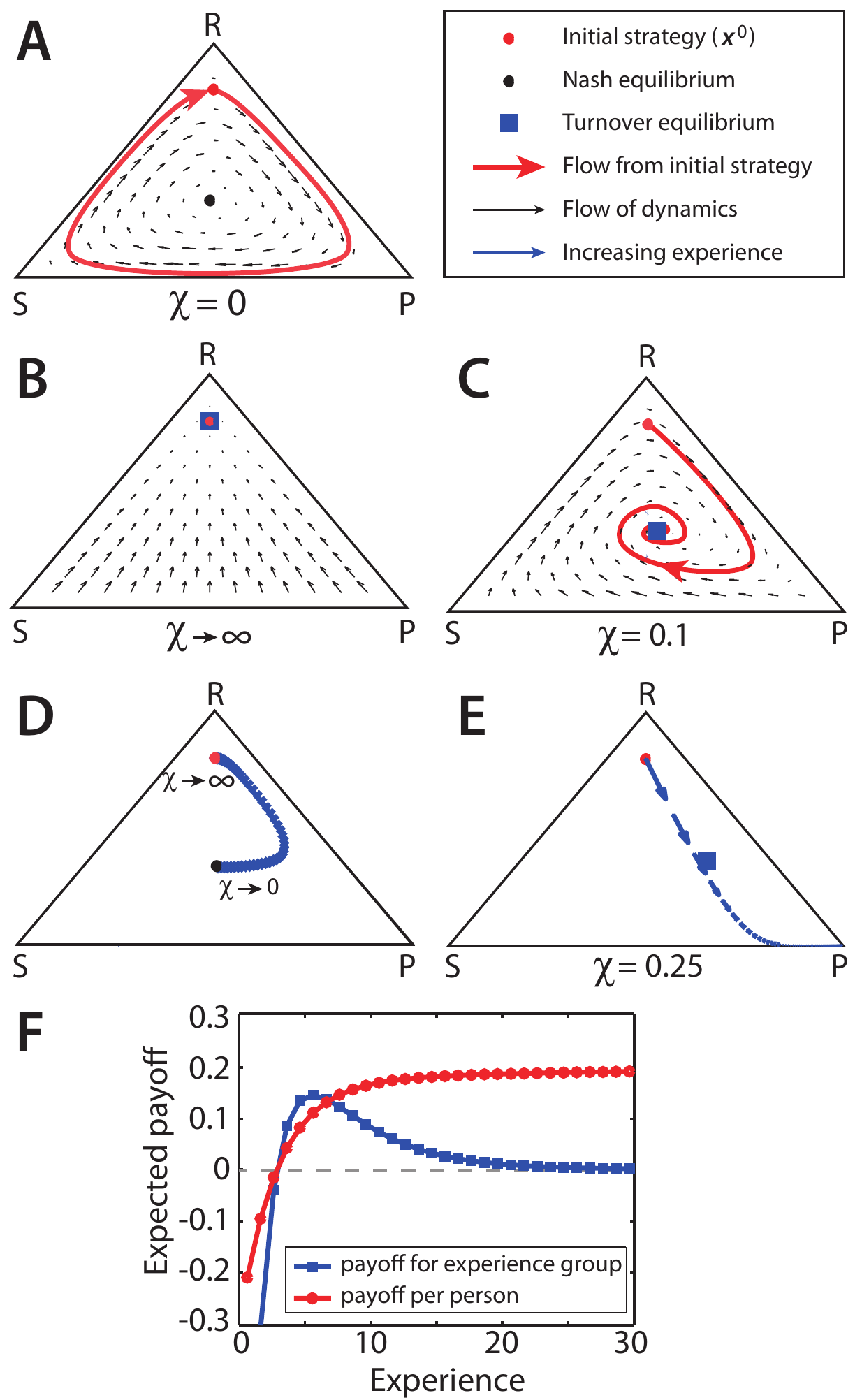}
	\caption{(Color online) Turnover of players in the
          rock-paper-scissors game 
with an initial strategy of $80 \%$ rock and $10 \%$ paper and scissors. 
          \textbf{A} For $\chi=0$, the average strategy oscillates in a fixed distance from the Nash equilibrium.
          \textbf{B} For infinite  turnover, only the initial strategy $\mathbf{x^0}$ is employed.
          \textbf{C} For intermediate turnovers, the average strategy converges to the turnover equilibrium.
          \textbf{D} The turnover equilibrium for different values of  $\chi$. The equilibrium is displaced towards paper, which  dominates the initial strategy.  
          \textbf{E} In the turnover equilibrium, inexperienced players quickly turn away from playing rock. Experienced players only play paper.
          \textbf{F} Inexperienced players have a large negative
          payoff, which monotonically increases as the players gain
          experience. However, since the number of experienced players
          falls off exponentially, most games are won by players of
          intermediate experience.  In this example, we chose a total
          population size $N=20$.
          \label{fig:RPS}}
	\end{center}
\end{figure}

In Fig. \ref{fig:RPS}A-D the time development of the average strategy
played is shown for an initial strategy $\mathbf{x}^0=(0.8,0.1,0.1)$
biased towards rock and for different values of $\chi$. Similar results 
are obtained using different initial strategies. In the absence
of agent turnover, the strategy of the population oscillates around
the Nash equilibrium \cite{szabo2007evolutionary, Reichenbach06}. Upon
introducing turnover, the initial overrepresentation of players
playing rock makes paper a rewarding strategy. As players change
strategy towards paper, scissors becomes more rewarding, and so
on.  In the limit of $\chi \to \infty$, all players remain na\"{\i}ve,
so the strategy of the population is equal to the initial
strategy.

For intermediate turnover rates, the system undergoes damped or
critically damped oscillations towards the turnover equilibrium, where
the gradual learning of the players staying in the game exactly
balances the exchange of experienced players with na\"{\i}ve
players. Thus, agent turnover stabilizes the game dynamics, in a
similar way as spatial organization of agents \cite{Kerr02,
  Reichenbach06, Juul12, Juul13}.  The location of the turnover equilibrium
depends non-trivially on the strategy of na\"{\i}ve players and on the
turnover rate, as shown in Fig. \ref{fig:RPS}D. When $\chi$ is
increased from $0$ to $\infty$, the turnover equilibrium shifts from
the Nash equilibrium to the strategy distribution of inexperienced
players along a curved line. This means that, for a generic value of
$\chi$, the turnover equilibrium is not trivially a linear combination
of the Nash equilibrium and the naive strategy.

In a traditional game of rock-paper-scissors, the time-averaged payoff
of all players is zero in any steady state situation. In the turnover
equilibrium, however, the average payoff of players increases
monotonically with their experience, such that new, na\"{\i}ve players
tend to lose to more experienced players.  Fig. \ref{fig:RPS}E shows
the average strategy $n_i(\tau,\infty)/N$ as a function of the
experience level $\tau$, where the turnover is $\chi = 0.25$ and the
na\"{\i}ve strategy is the same as before . The strategy with the
highest payoff is paper, which is the only strategy that very
experienced players employ. However, due to the turnover of agents,
the number of experienced players falls off exponentially. This causes
a radical qualitative difference between the payoff of a given player
having an experience level $\tau$ and the total payoff of players in a
given experience level. Denoting the turnover equilibrium by
$\mathbf{x}^*$, the total payoff collected by players with experience
$\tau$ is given by $\sum_i n_i(\tau,\infty)\pi_i(\mathbf{x}^*)$, while
the payoff of a single player with this experience is $\sum_i
n_i(\tau,\infty)\pi_i(\mathbf{x}^*)/(\sum_{i}
n_i(\tau,\infty))...$. The two different payoffs are shown in
Fig. \ref{fig:RPS}F as a function of the experience level $\tau$ for a
population of 20 players. The figure demonstrates the counterintuitive
fact that, while the payoff of each single player increase
monotonically with the experience, the majority of the games are won
by players with intermediate experience levels.

%%%%%%%%%%%%%%%%%%%%%%%%%%%%%%%%%%%%%%%%%%%%%%%%%%%%%%%%%%%%
\section{Two-agent two-action games} 
%%%%%%%%%%%%%%%%%%%%%%%%%%%%%%%%%%%%%%%%%%%%%%%%%%%%%%%%%%%%

We now turn to the broad class of games, where two populations can
both choose between two strategies. Let us define the mixed strategies
as $(x , 1-x)$ for the first population and $(y , 1-y)$ for the second
population. The payoffs obtained by the two populations depend on
their joint actions \cite{Hofbauer2003} and are traditionally encoded
in payoff matrices $A$ and $B$:
\begin{eqnarray} 
\begin{pmatrix} \pi_1^x \\ \pi_2^x \end{pmatrix} 
	= \begin{pmatrix} a_{11} & a_{12} \\ a_{21} & a_{22} \end{pmatrix} 	\begin{pmatrix} y \\ 1 - y \end{pmatrix}   \label{payoffX} \\ 
\begin{pmatrix} \pi_1^y \\ \pi_2^y \end{pmatrix}
	= \begin{pmatrix} b_{11} & b_{12} \\ b_{21} & b_{22} \end{pmatrix} 	\begin{pmatrix} x \\ 1 - x \end{pmatrix} ,  \label{payoffY}
\end{eqnarray}
where $\pi_1^x$ is the payoff of the first strategy for the first
population, and so forth.  The two populations are learning
concurrently and their turnover rates $\chi_x$ and $\chi_y$ can in
principle be different. The learning dynamics can be obtained from
Eq.~\eqref{repl_turnov}. 

\begin{eqnarray}
	\frac{d}{dt} x = x (1-x)(\pi_1^x - \pi_2^x)+\chi_x(x^0-x) \label{eq:repx1} \\
	\frac{d}{dt} y = y (1-y)(\pi_1^y - \pi_2^y)+\chi_y(y^0-y) \label{eq:repy1}
\end{eqnarray}
where the equations for the second strategies can be simply obtained
from the normalization conditions.  In the absence of turnover, this
system is known as a bi-matrix replicator equation
\cite{Hofbauer1996,Hofbauer2003}. In the turnover equilibrium, which
we denote as $(x , y) = (x^*, y^*)$, the time derivatives are equal to
zero. Upon substituting the expressions \eqref{payoffX} and \eqref{payoffY}
for the payoffs leads to the equations
\begin{eqnarray}
\alpha y^* + a_{12}-a_{22} = \chi_x \frac{x^*-x^0}{x^*(1-x^*)} , 
\label{turnEqui1} \\
\beta x^* + b_{12}-b_{22} = \chi_y \frac{y^*-y^0}{y^*(1-y^*)} , 
\label{turnEqui2}
\end{eqnarray}
where we have introduced 
\begin{align} 
	\alpha &= a_{11}+a_{22}-a_{21}-a_{12} , \label{alpha} \\
	\beta &= b_{11}+b_{22}-b_{21}-b_{12} . \label{beta} 
\end{align}
The number of solutions to equations \eqref{turnEqui1} and
\eqref{turnEqui2} depends on the sign of the product $\alpha \beta$.
When $\alpha \beta < 0$, there is always one mixed turnover
equilibrium.  When $\alpha \beta > 0$, the number of turnover
equilibria can increase at critical values of the turnover
parameter.
The derivation of these result can be found in appendix
\ref{app:2actionEquilibria}. In the following, we provide examples for
two well-known two-action games belonging to the two categories
$\alpha \beta < 0$ and $\alpha \beta > 0$: the game of matching
pennies and a coordination game, respectively.

%%%%%%%%%%%%%%%%%%%%%%%%%%%%%%%%%%%%%%%%%%%%%%%%%%%%%%%%%%%%
\subsection{Matching pennies}
%%%%%%%%%%%%%%%%%%%%%%%%%%%%%%%%%%%%%%%%%%%%%%%%%%%%%%%%%%%%
A simple example of $\alpha \beta < 0$ is the zero sum game of
matching pennies. Two players secretly turn a penny each to heads or
tails and reveal the coins simultaneously. If the pennies match,
player one receives a reward $r$ from player two. If they do not,
player one pays the reward to player two. This game is characterized
by the payoff matrices

\begin{align}
	A = \begin{pmatrix} r & -r \\ -r & r \end{pmatrix}
	\quad
	B = \begin{pmatrix} -r & r \\ r & -r \end{pmatrix} .
\end{align}

\begin{figure}[tb]
	\includegraphics[width=8.4cm]{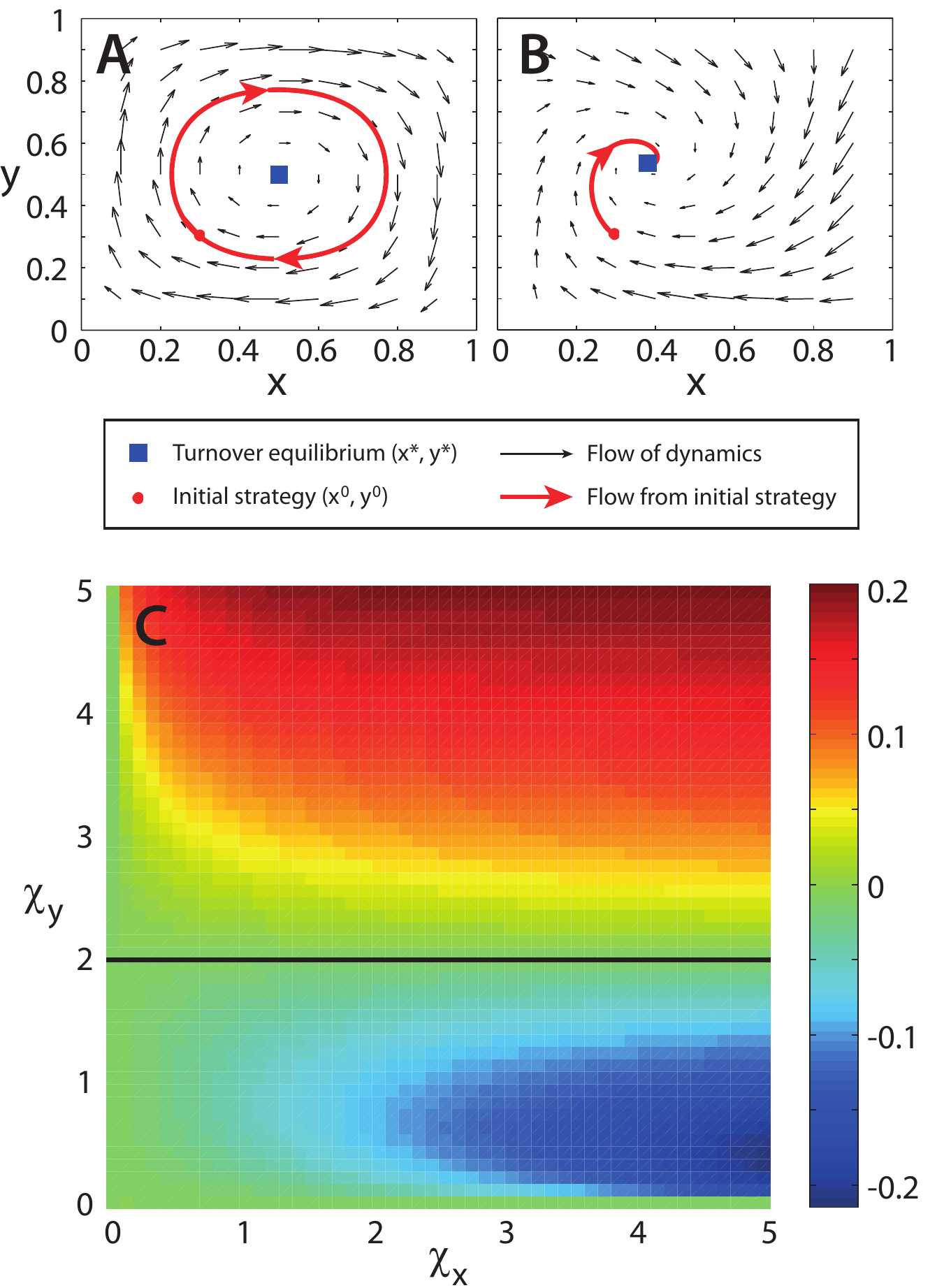}
  \caption{(Color online) Dynamics of the matching pennies game with $r=2$ and initial strategy $x^0 = y^0 = 0.3$. Other choices of parameter values yield similar results. 
  \textbf{A} For $\chi=0$, the average strategy oscillates at a fixed distance from the Nash equilibrium. 
  \textbf{B} When a turnover is introduced, the average strategy converges to the turnover equilibrium. Here $\chi_x = \chi_y = 1$. 
  \textbf{C} Expected payoff for population one in the turnover equilibrium for varying turnovers. Above a critical value of $\chi_y = 2$, the first population will win most games.}
      \label{fig:pennies}
\end{figure}

The learning dynamics \eqref{eq:repx1} and \eqref{eq:repy1} with no turnover leads to a marginally stable Nash equilibrium surrounded by concentric closed orbits (see Fig.~\ref{fig:pennies}A). When a positive turnover of players is introduced, the Nash equilibrium is perturbed to a stable turnover equilibrium $(x^*, y^*)$, as displayed in Fig. \ref{fig:pennies}B. Inserting the payoff matrices into the conditions for turnover equilibrium \eqref{turnEqui1} and \eqref{turnEqui2} gives
\begin{align}
	y^* = \frac 12 + \frac{\chi_x}{4r} \frac{x^*-x^0}{x^*(1-x^*)} , \label{MatchTurnEqui1} \\
	x^* = \frac 12 + \frac{\chi_y}{4r} \frac{y^*-y^0}{y^*(1-y^*)} , \label{MatchTurnEqui2}   
\end{align}

Matching pennies is a zero sum game. The expected
payoff of population one is positive if the two populations have
a tendency of both playing heads or both playing tails, while it is
zero if and only if one of the populations chooses their strategy
randomly. From Eq. \eqref{MatchTurnEqui1} and \eqref{MatchTurnEqui2}
we see that this happens in the turnover equilibrium if
$(x^*,y^*)=(1/2,y^0)$ or $(x^*,y^*)=(x^0,1/2)$. For a given set of
initial strategies of the populations, this corresponds to the critical
turnover rates
\begin{align}
	\chi_{x,c} = - r \frac{1-2 y^0}{1-2 x^0}   \label{MatchEquiChiX} \\
	\chi_{y,c} = r \frac{1- 2 x^0}{1-2 y^0} . \label{MatchEquiChiY}   
\end{align}
Equation \eqref{MatchEquiChiX} can only be satisfied if
population two wins most games when the initial strategies are
employed. In this case, the expected payoff of population one is
negative if it has a high turnover, since its equilibrium strategy
is close to its initial strategy. If the turnover is decreased 
below the critical value \eqref{MatchEquiChiX}, the first population
starts winning more games than it loses regardless of the turnover
of population two. Even if population two has an even lower turnover,
and thus has an initially superior strategy and more time to gain
experience, the equilibrium state remains advantageous for population
one.

Likewise, equation \eqref{MatchEquiChiY} only gives a positive
critical turnover if population one wins most games when the
initial strategies are employed. Here, the payoff for population two
goes from being positive to negative when its turnover is increased
past the critical value given by \eqref{MatchEquiChiY}, regardless of
the turnover of the first population. In Fig. \ref{fig:pennies}C, the
average payoff for population one is shown for varying turnovers of
both populations and the initial strategies $x^0 = y^0 = 0.3$. In
this case, if the turnover of population two is larger than
$\chi_{y,c} = 2$, population one wins most games.

%%%%%%%%%%%%%%%%%%%%%%%%%%%%%%%%%%%%%%%%%%%%%%%%%%%%%%%%%%%%
\subsection{Coordination game}
\begin{figure}[tb]
\centering
\includegraphics[width=8.4cm]{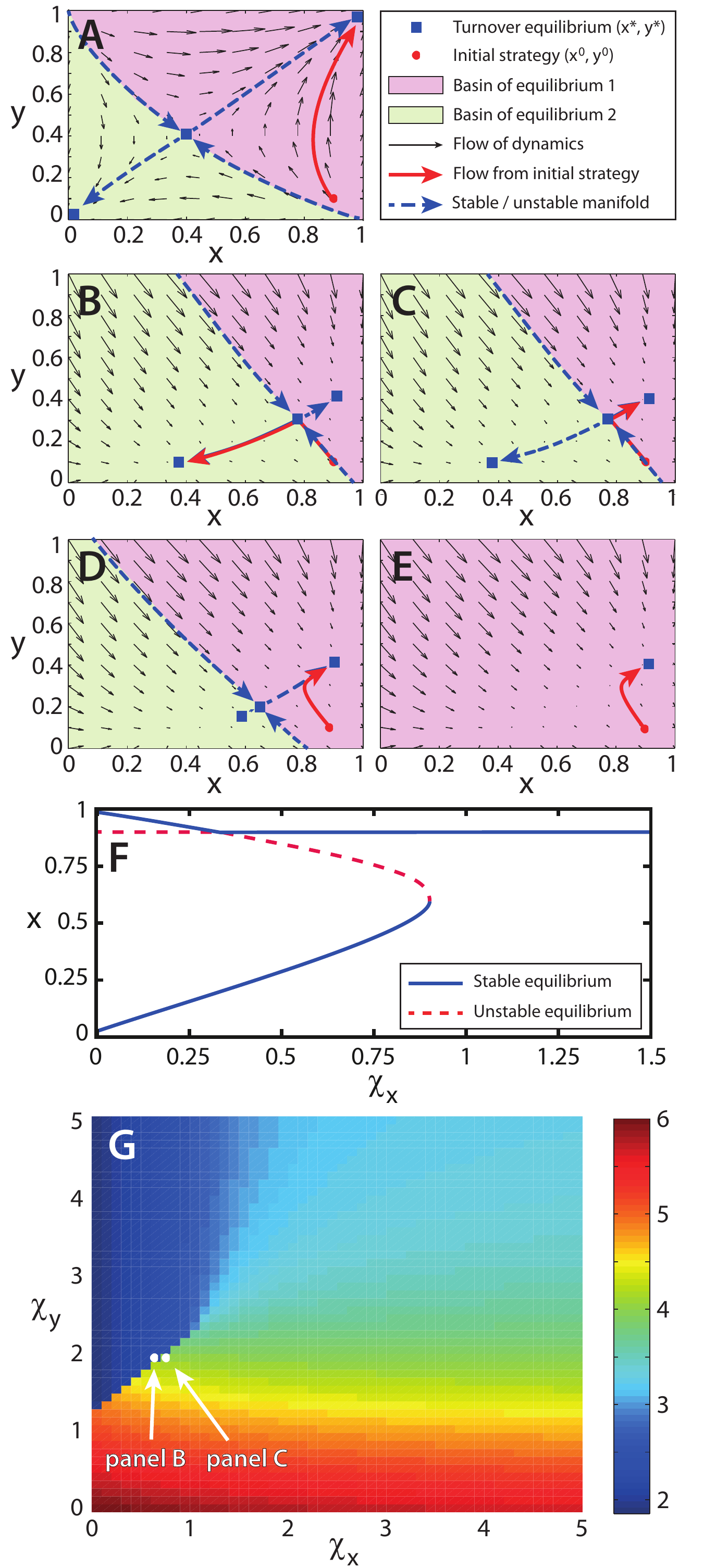}
\caption{(Color online) Dynamics of a coordination game with
  initial strategy $(x^0,y^0) = (0.9,0.1)$.  \textbf{A} Without
  turnover, the game has one saddle point Nash equilibrium with stable
  manifolds that separate the basins of attraction of two pure Nash
  equilibria.  \textbf{B-C} When the initial strategy goes from one
  basin of attraction to another, the resulting equilibrium state
  changes discontinuously. Here $\chi_y = 2$ and $\chi_x = 0.7$ and
  $0.72$, respectively.  \textbf{D-E} At a critical set of turnover
  rates two turnover equilibria annihilates in a saddle node
  bifurcation. Here $\chi_y = 2$ and $\chi_x = 0.91$ and $0.92$,
  respectively.  \textbf{F} Bifurcation diagram for turnover equilibria. The point $(x,y) = (0.4 , 0.9)$ is always an equilibrium. We observe a transcritical bifurcation at $\chi_x \approx 0.4$ and the saddle node bifurcation from panels D-E at $\chi_x \approx 0.9$.
  \textbf{G} Expected payoff of population one in the
  turnover equilibrium for varying turnovers. The dramatic change in
  payoffs between panel B and C can clearly be seen.
  \label{fig:coordination}}
\end{figure}

%%%%%%%%%%%%%%%%%%%%%%%%%%%%%%%%%%%%%%%%%%%%%%%%%%%%%%%%%%%%
When the payoff matrices of the two populations equal each other,
$A=B$, one must necessarily have $\alpha \beta \ge 0$. One example of
this is a coordination game, where the two players strive to play the
same strategy. We consider the payoff matrices

\begin{align}
	A = B = \begin{pmatrix} 6 & 0\\ 3 & 2 \end{pmatrix} ,
\end{align}
where $\alpha = \beta = 5$. Coordination games with different payoff matrices 
will have similar dynamical properties. Without player turnover, this game has
two pure Nash equilibria where the populations employ the same
strategy. In addition, there is a mixed Nash equilibrium at
$(x^*,y^*)=(2/5,2/5)$ with one stable and one unstable manifold. The
stable manifold constitutes the boundary between the basins of
attraction of the two pure Nash equilibria (see
Fig. \ref{fig:coordination}A). For sufficiently small turnovers, there
are three turnover equilibria close to these three Nash
equilibria. The initial strategies $(x^0 , y^0)$ determine which
equilibrium state the system goes to.

Let us consider a situation of strong initial disagreement $(x^0 ,
y^0) = (0.9,0.1)$, where inexperienced players of population one and
two tend to employ the first and second strategy, respectively. We fix
$\chi_y=2$, while varying $\chi_x$ as a control parameter. In this
case, the point $(x,y) = (0.4 , 0.9)$ is always an equilibrium of the
dynamics. At a critical value, $\chi_x \approx 0.71$, the boundary
between the basins of attractions passes the initial strategy $(x^0 ,
y^0)$ (see Fig. \ref{fig:coordination}B-C), and the steady state
changes discontinuously from being dominated by strategy two to
strategy one. At the same point, the payoff of both populations
increases drastically.

Upon increasing the turnover of the first population even further, the
basin of attraction of strategy two suddenly disappears at $\chi_x
\approx 0.91$. For this value, the saddle node equilibrium annihilates
with the turnover equilibrium of strategy two (see
Fig. \ref{fig:coordination}D-E). For larger values of $\chi_x$, there
is only one turnover equilibrium. In this case the number of turnover
equilibria changes through a saddle node bifurcation, but for other
payoff matrices A, B, a pitchfork bifurcation could occur (see
appendix \ref{app:2actionEquilibria}).

A full bifurcation diagram is shown in
Fig. \ref{fig:coordination}F, and the expected payoff of population
one is shown in Fig. \ref{fig:coordination}G as a function of the
turnover rates. The abrupt change in payoff at a critical set of
turnover rates can clearly be seen. The expected payoff of population
two follows a similar pattern. In general, a low turnover of either
population results in higher payoffs of both populations, since a
larger proportion of the players has enough experience to mainly bid
on the dominating strategy. However, increasing the turnover may
change which equilibrium the system goes to, leading to an increase in
the payoff of both players.

%%%%%%%%%%%%%%%%%%%%%%%%%%%%%%%%%%%%%%%%%%%%%%%%%%%%%%%%%%%%
\section{Lowest Unique Bid Auctions}\label{aucsec}
%%%%%%%%%%%%%%%%%%%%%%%%%%%%%%%%%%%%%%%%%%%%%%%%%%%%%%%%%%%%
Lowest unique bid auctions have become popular online games, and have
recently attracted attention from the scientific community \cite{Baek,
  Ostling, Pigolotti}. In this game, $N$ players pay an entrance fee
to independently bid on an item, \emph{e.g.} a car. The winner of the
auction is the player with the lowest bid not also plaved by another
player. The strategies available to each player are the possible
integer bids $i = \{1, 2, 3 \ldots \}$.

If a population of players place their bids stochastically according
to the distribution $x_i$, the average payoff of each strategy is, in
the large $N$ limit, proportional to \cite{Pigolotti}

\begin{equation}\label{payoffAuction}
  \pi_i = e^{-N x_i} \prod\limits_{j=1}^{i-1} (1-N x_i e^{-N x_i}).
\end{equation}

The Nash equilibrium corresponding to the payoffs of Eq. 
\ref{payoffAuction} was compared with data from 
the auctions website {\it auctionair.co.uk} 
in \cite{Pigolotti}. An excellent agreement between the
Nash equilibrium and the data was found for low to intermediate values of
$N$. In larger auctions, the bidding distribution departed from the
Nash equilibrium and was more resemblant of an exponential distribution
\begin{equation}\label{x0Auction}
  x_i^0 = \frac{ e^{-\beta i}}{ \sum_i e^{-\beta i}} .
\end{equation}
where the fitted value of the exponential constant was $\beta \approx
0.02$. On general theoretical grounds, one can argue that the
exponential bidding strategy is the expected prior distribution for
inexperienced players, with limited information about the game, that
want to avoid the cost of a large bid \cite{Pigolotti, Goeree}.

Using \eqref{payoffAuction} and \eqref{x0Auction}, we test here the
hypothesis that the apparent change of behavior from small to large
auction sizes is caused by agent turnover. Note that changing the
turnover rate $\chi$ in \eqref{repl_turnov} is equivalent to
multiplying the payoff function \eqref{payoffAuction} with a constant
and rescaling time. As it is also problematic to infer a natural time
scale of adaptation from the data, we leave the value of $\chi$ as a fitting
parameter.

\begin{figure}[]
  \includegraphics[width=8.4cm]{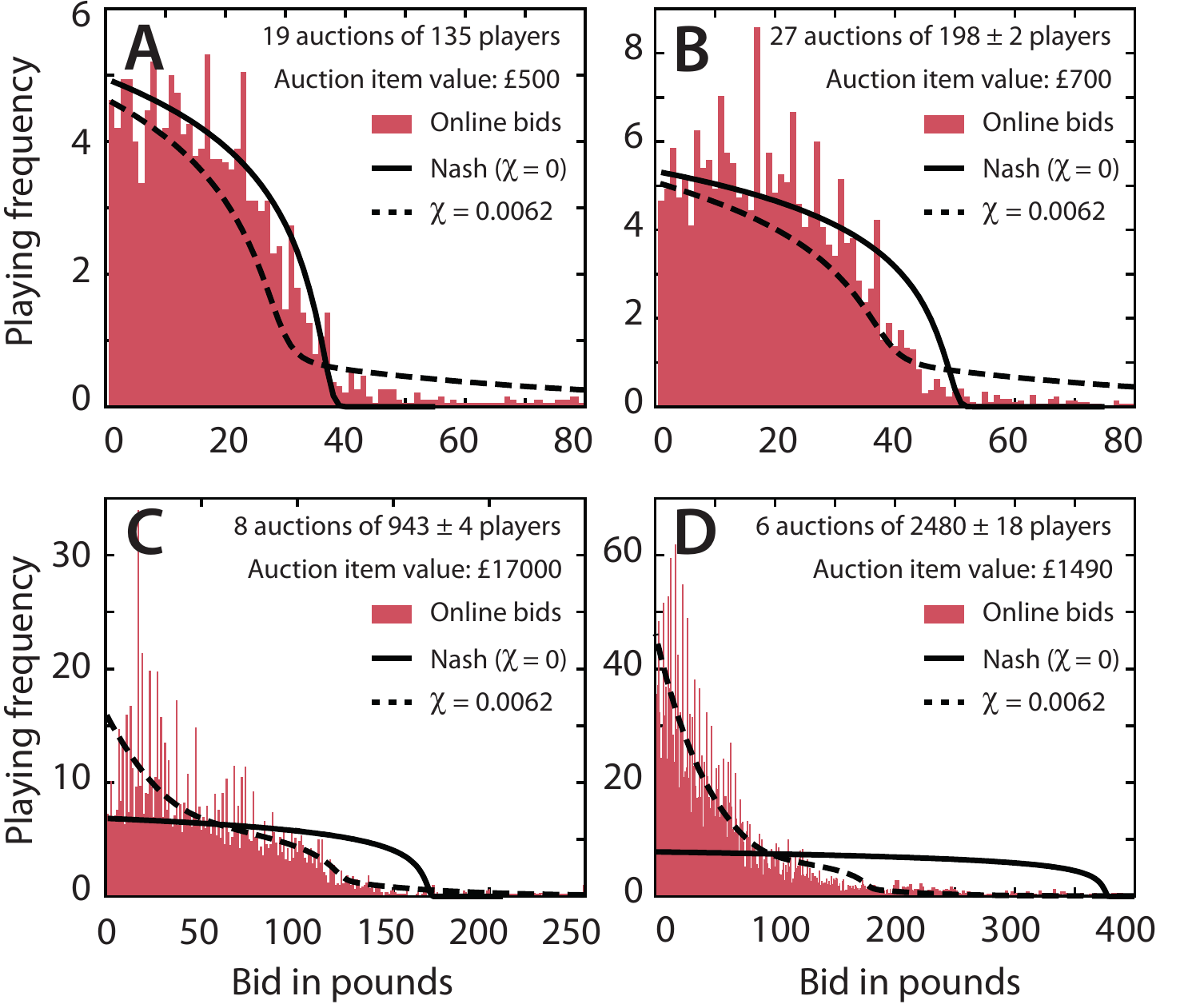}
  \caption{(Color online) Bidding frequencies in online lowest unique
    bid auctions compared to the theoretical Nash
    equilibrium %, which is reached if $\chi = 0$,
    and the turnover equilibrium for a constant $\chi =
    0.0062$. Different panels show auctions with items of different
    value and different numbers of players. For small auction sizes
    both the Nash and the turnover equilibrium fit the empirical
    bidding distributions. In larger auctions the turnover equilibrium
    fits the data much better than the Nash equilibrium. This is due
    to a long adaptation time for players in large
    auctions. \label{fig:auction}}
\end{figure}

Fig. \ref{fig:auction} compares empirical data from 60 online
auctions with varying number of players to the respective Nash
equilibria and turnover equilibria for a least square fitted value of $\chi=0.0062$. We define the squared distance $d$ between the empirical data and the turnover and Nash equilibria as
\begin{align}
d_{turn}=N^2 \sum_i (x_i^{emp}-x_i^*)^2 \label{eq:distTurn} \\
d_{Nash}=N^2 \sum_i (x_i^{emp}-x_i^{Nash})^2  \label{eq:distNash}
\end{align}
where $\mathbf{x}^{emp}$ is obtained from the frequencies, $\mathbf{x}^{Nash}$ is the theoretical Nash equilibrium, and the sum runs over all bids and all auctions. Using these definitions we get $d_{turn} = 89084$ and $d_{Nash} = 253526$. In \cite{Pigolotti} it is shown that, if the bids were randomly drawn from the theoretical distribution, the squared distance is expected to be equal to the combined number of bidders $d_{exp} = \sum N = 30335$. This shows that the turnover equilibria describe the bidding distributions much better than the Nash equilibria, introducing just a single free parameter.

In small auctions, both the Nash equilibrium and the turnover
equilibrium fit the data well, but the latter also captures the fat
tail of the empirical distribution. In larger auctions, the Nash
equilibrium fails to predict the observed distribution of bids, while
the turnover equilibrium fits the data remarkably well.

It should be reminded that, while the exponential strategy
distribution for large $N$ is plugged into the model empirically as a
prior, it is still remarkable that, for the same value of $\chi$, the
resulting equilibrium is relatively close to the prior for large $N$
and closer to the Nash equilibrium for smaller values of $N$. The
reason stems from the growth of the adaptation time of the replicator
dynamics with the auction size $N$ (see \cite{Pigolotti}). The
consequence if, even at equal turnover rate $\chi$, players have
sufficient time to adapt for small auction size but not for larger
auctions. Clearly, one could get an even better fit by letting $\chi$
vary as a function of $N$, as one could have different kinds of
players and therefore different turnovers for different auction types,
as exemplified in Appendix \ref{app_auc}. We do not present a
systematic study of the dependence of $\chi$ on $N$, as it is beyond
the scope of this paper. Our main point here is to show that, even in
the simple case of a fixed $\chi$, the replicator equation with
turnover is useful to model a real-life game and can help
conceptualizing non-trivial discrepancies between observed strategy
distributions and Nash equilibria. Finally, in Appendix \ref{app_auc}
we show that, in the case of larger auctions, the distance from the
Nash equilibrium does not seem to depend on time. This fact further supports the
presence of a steady effect, like the turnover mechanism proposed
here, preventing the population to reach the Nash equilibrium.

%%%%%%%%%%%%%%%%%%%%%%%%%%%%%%%%%%%%%%%%%%%%%%%%%%%%%%%%%%%%
\section{Discussion and conclusion}
%%%%%%%%%%%%%%%%%%%%%%%%%%%%%%%%%%%%%%%%%%%%%%%%%%%%%%%%%%%%

In this paper we have extended the theoretical framework of the
replicator equation by considering the effect of agent turnover, a
phenomenon that is in principle present in many game theoretical
contexts.  Our model provides a simple and compact description of the
dynamics of the average strategies in the population, and at the same
time allows for dissecting the behavior of players having different
experience levels. The study of experience-dependent payoffs can bring
to counterintuitive results, as we demonstrated in the simple case of
a Rock-Paper-Scissor game. When turnover is introduced, the
game dynamics become richer. We have shown that it is possible to
encounter bifurcations of turnover equilibria, and that the 
equilibrium strategy can change discontinuously with the turnover rate.

The simplicity and flexibility of the approach proposed in this paper
can be of use to analyze several games of practical importance.  For
example, we have shown here how turnover of agents can help
explaining adaptation in online games, where experienced players
may exploit the na\"{\i}ve strategies of new players. Our comparison
with a dataset of bidding distributions in online lowest unique bid
auctions shows that our model is able to describe the empirical steady
state distributions far better than the Nash equilibrium. We expect
that similar results can be obtained by studying data from other
games.

The concepts presented here suggest an interesting analogy with the
idea of \emph{cognitive hierarchy} \cite{costamesa,camerer,wright} in
human learning, where players are classified according to the number
of reasoning steps they are able to make. In this scheme,
non-reasoning players are reminiscent of inexperienced players in the
present model, and players with more reasoning steps can be compared
more to experienced players. In this perspective, the model proposed
here provides a framework in which a cognitive hierarchy emerges
naturally from the dynamical equilibrium between adaptation and
turnover. Indeed, a cognitive hierarchy model has previously been
applied to describe the bidding distributions lowest unique bid
auctions \cite{Ostling}. In the authors' of Ref. \cite{Ostling} own
words, \emph{``the cognitive hierarchy model $\ldots$ should be viewed
  as a potential stepping stone to an investigation using a formal
  learning model''}. Our work is one proposal of such a learning
model.

It is known that a fraction of non-rational, influenceable players may
change the game dynamics considerably \cite{halpern1998debating,
  nowak2000fairness, lorentziadis2012optimal}.  In a recent study
\cite{couzin2011uninformed}, a population of fish were made up of
three subgroups: a large group of fish with a small preference for
going to one place in the aquarium, a smaller group with a strong
preference for another place, and a group of untrained fish with no
prior preference. The study showed, both through simulations and
experiment, that there exists a critical size of the untrained group
above which the entire population goes to the place prefered by group
one, and below which the minority of group two dictates where the
population goes. This resembles our results for two-agent two-action
coordination games, where a critical turnover of agents changes the
turnover equilibrium from being dominated by one strategy to the
other.

In our model we assume a constant turnover of agents in time, which
leaves one free fitting parameter. However, changing the turnover is
mathematically equivalent to scaling the expected payoff of all
strategies together with a time scaling. Since the absolute value of
the payoff is often treated as a free parameter when analyzing
experiments of learning dynamics, this does not increase in general the
number of fitting parameters in such an analysis. Furthermore, if the
absolute values of the payoffs are known, one can in principle measure the
turnover rate of agents and avoid having a free fitting parameter.

In conclusion, we have shown how a steady turnover of agents, which is
present in most real-life games, can change the qualitative dynamics
of a game. The simplicity and generality of the framework presented
here makes it a natural candidate to describe adaptation in a
population participating in a game and subject to turnover.

\begin{acknowledgments}
We thank A. Galstyan and C. Strandkvist for comments on a preliminary version of the manuscript. A. Kianercy research was supported in part by the National Science Foundation under grant No. 0916534 and the US AFOSR MURI grant No. FA9550-10-1-0569.
\end{acknowledgments}

\appendix

%%%%%%%%%%%%%%%%%%%%%%%%%%%%%%%%%%%%%%%%%%%%%%
\section{Number of turnover equilibria in two-agent two-action games} \label{app:2actionEquilibria}
%%%%%%%%%%%%%%%%%%%%%%%%%%%%%%%%%%%%%%%%%%%%%%
In this appendix, we show that the number of possible turnover
equilibria in two-agent two-action games is dependent on the sign of
the product $\alpha \beta$, where $\alpha$ and $\beta$ are given by
\eqref{alpha} and \eqref{beta}.

The conditions for turnover equilibrium are given by \eqref{turnEqui1}
and \eqref{turnEqui2}. Expressing $y^*$ as a function of $x^*$, this
can be written as

\begin{eqnarray}
	y^*(x^*)  = \frac{\chi_x}{\alpha} \frac{x^*-x^0}{x^*(1-x^*)} - \frac{a_{12}-a_{22}}{\alpha}, \label{yStar} \\
	\beta x^* + b_{12}-b_{22} = \chi_y \frac{y^*(x^*)-y^0}{y^*(x^*)(1-y^*(x^*))} \equiv g(x^*) , \label{newCondition}
\end{eqnarray}
where we defined the composite function $g(x^*)$. Notice that
$\lim_{x^*\rightarrow 0} g(x^*) = \lim_{x^*\rightarrow 1} g(x^*) =0$
and that $g(x^*)$ has two singularities for values of $x^*$ such that
$y^*=0$ and $y^*=1$. Since $0\le y^*\le 1$, we will focus on the
interval of $g(x^*)$ between the two singularities.

\begin{figure}[tb]
	\centering
	\includegraphics[width=8.4cm]{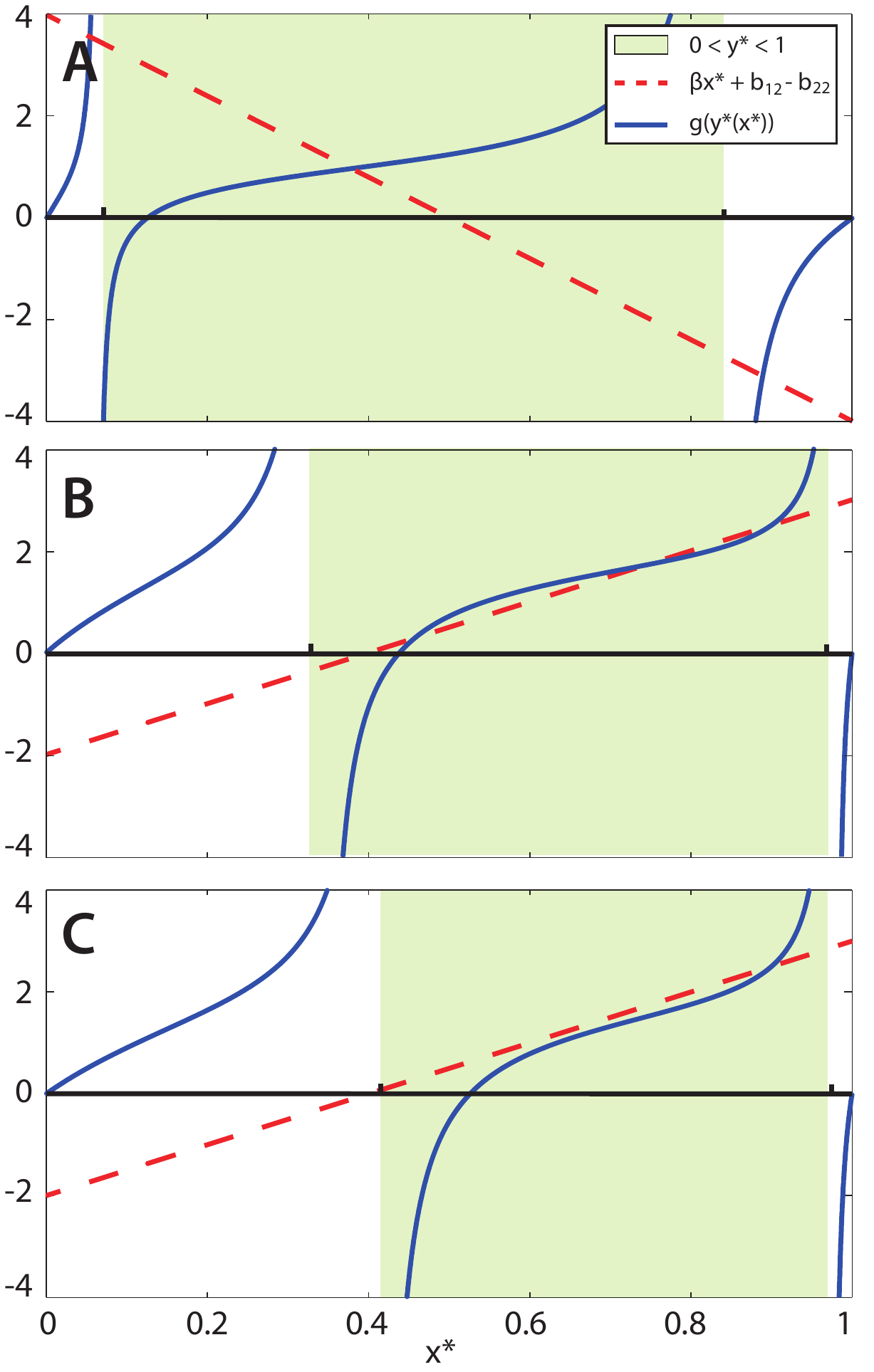}
\caption{(Color online) The number of turnover equilibria in a two-action two-agent
  game as determined by a graphical solution of equation \ref{newCondition}. 
\textbf{A} When $\alpha \beta <0$ the two functions have opposite slopes and hence intersect exactly once in the relevant interval $0<y^*<1$. Here the matching pennies game is illustrated with the parameters of Fig. \ref{fig:pennies}B.
\textbf{B-C} When $\alpha \beta >0$ the slope of the functions have
the same sign, and new turnover equilibria can appear through
bifurcations. Here the bifurcation in the coordination game of
Fig. \ref{fig:bifurcation}B-C is shown, except that $\chi_x$ is changed from to $0.8$ to $1$ between panels B-C.
\label{fig:bifurcation}}
\end {figure}

Let us investigate the derivative $\partial g/\partial
  x^*$. From Eq. \eqref{yStar} and \eqref{newCondition} we get
\begin{eqnarray}
\frac{\partial y^*}{\partial x^*} &= \frac{\chi_x}{\alpha} 
\frac{(x^*-x^0)^2+ x^0- {x^0}^2}{(1-x^*)^2 {x^*}^2} \label{yp_EQ} \\
\frac{\partial g}{\partial x^*} &= \chi_y \frac{(y^*-y^0)^2 + 
y^0-{y^0}^2}{(1-y^*)^2 {y^*}^2} \frac{\partial y^*}{\partial x^*} \label{gp_EQ} .
\end{eqnarray}

Since $x^0$ and $y^0$ are probabilities, we always have $x^0 >
{x^0}^2$ and $y^0 > {y^0}^2$. The fractions in Eq. \eqref{yp_EQ} and
\eqref{gp_EQ} are therefore positive, so the derivative
$\partial g/\partial x^*$ must have the same sign as
$\alpha$. This means that $g(x^*)$ is either increasing or decreasing
monotonically, depending on the sign of $\alpha$.

If $\alpha \beta <0$, the line $\beta x^* + b_{12}-b_{22}$ and the
function $g(x^*)$ have opposite slopes, and therefore intersect
exactly once. It follows from \eqref{newCondition} that there is 
always exactly one turnover equilibrium. In
Fig. \ref{fig:bifurcation}A, this is illustrated for the matching
pennies game with the same parameters as Fig. \ref{fig:pennies}B.

If $\alpha \beta > 0$, the line $\beta x^* + b_{12}-b_{22}$ and the
function $g(x^*)$ are either both increasing or both
decreasing. Therefore, they can in principle intersect any odd number
of times. Hence, it is possible to have multiple turnover equilibria,
and these can appear or annihilate in pairs through either saddle
point bifurcations or pitchfork
bifurcations. Fig. \ref{fig:bifurcation}B-C shows a saddle node
bifurcation in the coordination game with the same parameters as
Fig. \ref{fig:coordination}D-E, except we have set $\chi_x$ equal to
$0.8$ and $1$, respectively, to show a clearer bifurcation.

%%%%%%%%%%%%%%%%%%%%%%%%%%%%%%%%%%%%%%%%%%%%%%
\section{Additional analysis of the lowest unique bid auction data}\label{app_auc}
\begin{figure}[]
  \includegraphics[width=8.4cm]{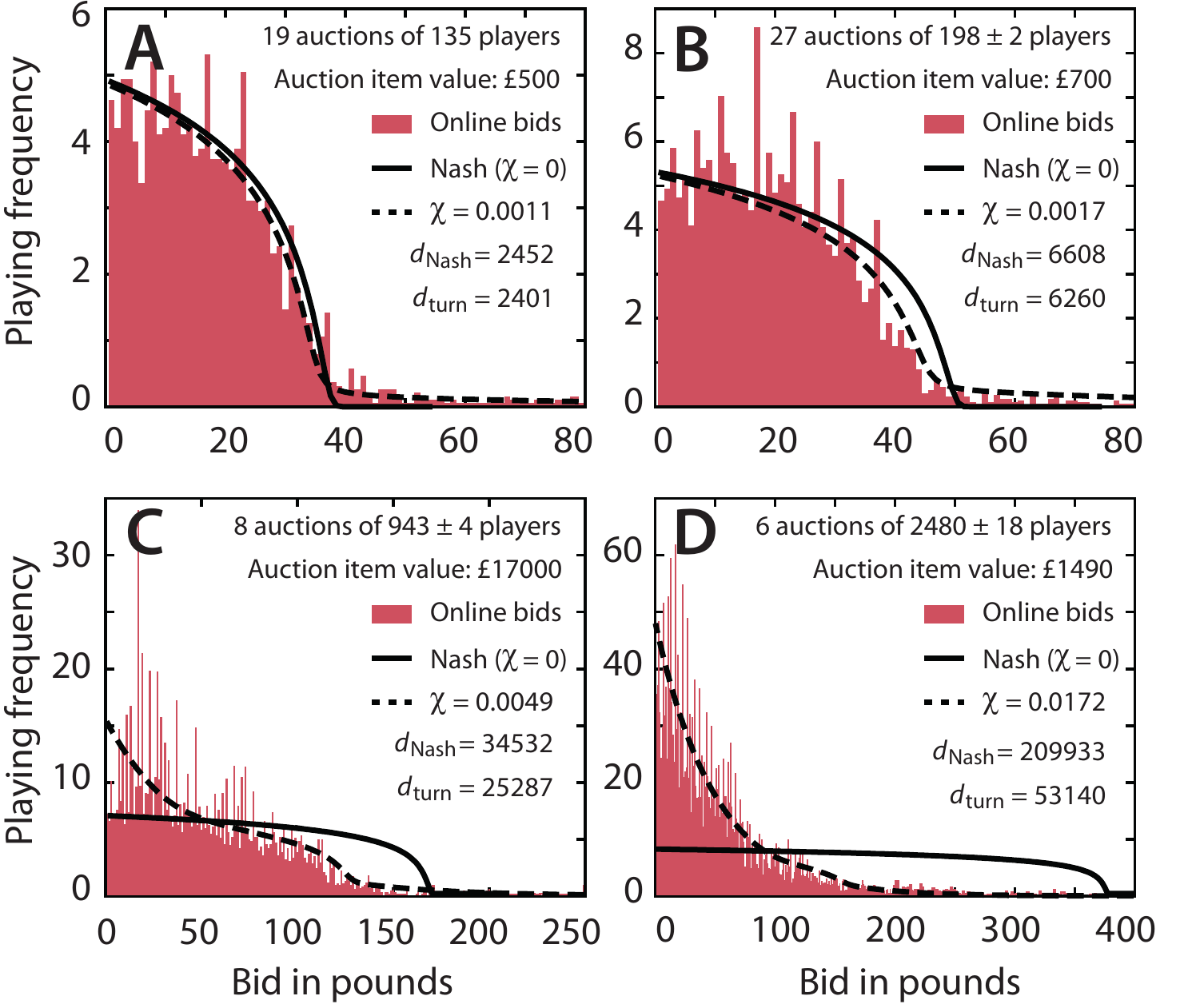}
  \caption{(Color online) Same data as Fig. \ref{fig:auction}, but
    this time the turnover parameter $\chi$ is fitted independently in
    each plot. The fitted value of $\chi$ and the total distances between the empirical bidding distributions and the turnover and Nash equilibria, respectively, are indicated in each legend.
    \label{fig:auction2}}
\end{figure}

%%%%%%%%%%%%%%%%%%%%%%%%%%%%%%%%%%%%%%%%%%%%%%
In this Appendix, we present additional results on the lowest unique bid
auction dataset used in Section \ref{aucsec}. Fig. \ref{fig:auction2}
shows how letting $\chi$ as an independent fitting parameter for each
auction size leads to very good fits of the bidding distributions in
all cases. The fitted values suggest a higher turnover rate for
larger auction sizes, which is reasonable considering for example that
larger auctions typically involve larger bidding fees. Furthermore, we see from the combined distances \eqref{eq:distTurn} and \eqref{eq:distNash} presented in Fig. \ref{fig:auction2} that we always have $d_{turn}<d_{Nash}$, so the turnover equilibria always fit the empirical data better than the Nash equilibria, but that this is most significant for large auction sizes. Both equilibria describe the bidding distributions well in small auctions.
Notice however
%that the relation is not strictly monotonic (fitted $\chi$ for $N=135$
%is slightly larger than for $N=198$). Moreover, it should be stressed
that the prior distribution $x_0$ has been obtained from the larger
auction sizes, so that estimates of $\chi$ in the large auction cases
can be affected by overfitting.

\begin{figure}[]
  \includegraphics[width=8.4cm]{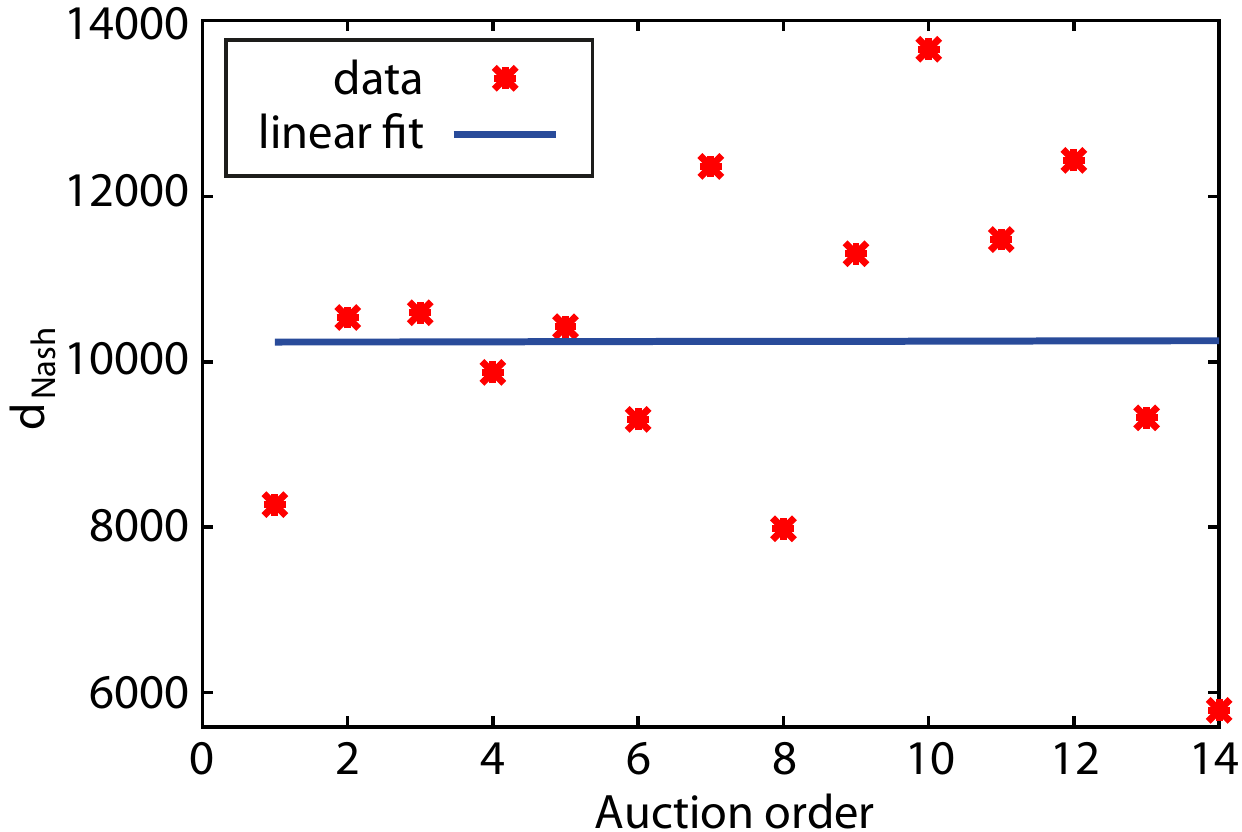}
  \caption{(Color online) Distance from the Nash equilibrium for
   14 auctions having $N=1398\pm74$, as a function of the time order
   of the auctions. The straight line is a fitted linear regression, with
   slope $m=1 \pm 140$.
    \label{fig:distance}}
\end{figure}

Finally, in Fig. \ref{fig:distance} we show the evolution of the
distance \eqref{eq:distNash} to the Nash equilibrium in series of auctions having similar
number of players $N=1398\pm74$.  The exact date of each auction was
not available, so that data are plotted simply as a function of the
time ordering of the auctions. 

It is easy to demonstrate that, if
players bid according to the Nash equilibrium, the expected value of
$d$ is $N$ (see \cite{Pigolotti}). A value of $d$ significantly larger
than $N$ signals a departure from the Nash equilibrium. Fitting the
data with a linear regression yields a slope $m=1 \pm 140$ that
clearly does not allow for refuting the null hypothesis of no temporal
trend. This suggests the presence of a mechanism keeping the
population at a finite distance from the Nash equilibrium, as the
turnover mechanism proposed in this paper.

%\bibliography{turnoverBibtex}

\end{document}